\documentstyle[10pt,epsfig,dp_delphititle,float,hangcaption,xspace,amssymb,
amsfonts,amsmath,amsthm,cite,graphicx]{dp_delphi}
\makeindex
\def\DpPaperGroup{PH-EP}
\def\DpPaperRef{2005-029}
\def\DpDate{30 June 2005}
\def\DpAuthors{DELPHI Collaboration}
\def\DpSubmit{(Accepted by Euro. Phys. Journ. C)}
\def\DpTitle{{ Study of double-tagged $\gamma\gamma$ events at LEPII}}

\newcommand {\pv} {$\pm$}
\newcommand {\wgg} {W_{\gamma^*\gamma^*}}
\newcommand {\agg} {$\gamma^*\gamma^*$ \hspace*{1mm}}
\newcommand {\sgg} {$\sigma_{\gamma^*\gamma^*}$ \hspace*{1mm}}
\newcommand {\see} {$\sigma_{ee}$ \hspace*{1mm}}
\newcommand {\dsee} {$d\sigma_{ee}/dY$ \hspace*{1mm}}
\newcommand {\dseed} {$(d\sigma_{ee}/dY)_{data}$ \hspace*{1mm}}
\newcommand {\dseemc} {$(d\sigma_{ee}/dY)_{MC}$ \hspace*{1mm}}
\newcommand {\dsgg} {$d\sigma_{\gamma^*\gamma^*}/dY$ \hspace*{1mm}}
\newcommand {\eehadt} {$e^+e^- \rightarrow hadrons$ \hspace*{1mm}}

\newcommand {\gghadt} {$\gamma^*\gamma^* \rightarrow hadrons$ \hspace*{1mm}}
\newcommand {\ggeemm} {$e^+e^- \rightarrow e^+e^- + \mu^+\mu^-$ \hspace*{1mm}}
\newcommand {\ggeeha} {$e^+e^- \rightarrow e^+e^- + hadrons$ \hspace*{1mm}}
\newcommand {\ggmmdt} {$\gamma^*\gamma^* \rightarrow \mu^+\mu^-$ \hspace*{1mm}}
\newcommand {\ggttdt} {$\gamma^*\gamma^* \rightarrow
\tau^+\tau^-$ \hspace*{1mm}}
\newcommand {\ggmmst} {$\gamma^*\gamma \rightarrow \mu^+\mu^-$ \hspace*{1mm}}
\newcommand {\gghast} {$\gamma^*\gamma \rightarrow hadrons$ \hspace*{1mm}}
\newcommand {\ggmmnt} {$\gamma\gamma \rightarrow \mu^+\mu^-$ \hspace*{1mm}}
\begin{document}
\makeatletter
\newcount\@tempcntc
\def\@citex[#1]#2{\if@filesw\immediate\write\@auxout{\string\citation{#2}}\fi
  \@tempcnta\z@\@tempcntb\m@ne\def\@citea{}\@cite{\@for\@citeb:=#2\do
    {\@ifundefined
       {b@\@citeb}{\@citeo\@tempcntb\m@ne\@citea\def\@citea{,}{\bf ?}\@warning
       {Citation `\@citeb' on page \thepage \space undefined}}%
    {\setbox\z@\hbox{\global\@tempcntc0\csname b@\@citeb\endcsname\relax}%
     \ifnum\@tempcntc=\z@ \@citeo\@tempcntb\m@ne
       \@citea\def\@citea{,}\hbox{\csname b@\@citeb\endcsname}%
     \else
      \advance\@tempcntb\@ne
      \ifnum\@tempcntb=\@tempcntc
      \else\advance\@tempcntb\m@ne\@citeo
      \@tempcnta\@tempcntc\@tempcntb\@tempcntc\fi\fi}}\@citeo}{#1}}
\def\@citeo{\ifnum\@tempcnta>\@tempcntb\else\@citea\def\@citea{,}%
  \ifnum\@tempcnta=\@tempcntb\the\@tempcnta\else
   {\advance\@tempcnta\@ne\ifnum\@tempcnta=\@tempcntb \else \def\@citea{--}\fi
    \advance\@tempcnta\m@ne\the\@tempcnta\@citea\the\@tempcntb}\fi\fi}
 
\makeatother
\begin{titlepage}
\pagenumbering{roman}
\CERNpreprint{\DpPaperGroup}{\DpPaperRef} 
\date{{\small\DpDate}} 
\title{\DpTitle} 
\address{\DpAuthors} 
\begin{shortabs} 
\noindent
Double-tagged interactions of photons with virtualities $Q^2$ between 
10 GeV$^2$ and 200 GeV$^2$ are studied with the data collected by 
DELPHI at LEPII from 1998 to 2000, corresponding 
to an integrated luminosity of 550 pb$^{-1}$.
The \ggmmdt data agree with QED predictions.
The cross-section of the reaction \gghadt
is measured and compared to the LO and NLO BFKL calculations.
\end{shortabs}
\vfill
\begin{center}
\DpSubmit \ \\ 
\end{center}
\vfill
\clearpage
\headsep 10.0pt
\addtolength{\textheight}{10mm}
\addtolength{\footskip}{-5mm}
\begingroup
%
\newcommand{\DpName}[2]{\hbox{#1$^{\ref{#2}}$},\hfill}
\newcommand{\DpNameTwo}[3]{\hbox{#1$^{\ref{#2},\ref{#3}}$},\hfill}
\newcommand{\DpNameThree}[4]{\hbox{#1$^{\ref{#2},\ref{#3},\ref{#4}}$},\hfill}
\newskip\Bigfill \Bigfill = 0pt plus 1000fill
\newcommand{\DpNameLast}[2]{\hbox{#1$^{\ref{#2}}$}\hspace{\Bigfill}}
%
\footnotesize
\noindent
\DpName{J.Abdallah}{LPNHE}
\DpName{P.Abreu}{LIP}
\DpName{W.Adam}{VIENNA}
\DpName{P.Adzic}{DEMOKRITOS}
\DpName{T.Albrecht}{KARLSRUHE}
\DpName{T.Alderweireld}{AIM}
\DpName{R.Alemany-Fernandez}{CERN}
\DpName{T.Allmendinger}{KARLSRUHE}
\DpName{P.P.Allport}{LIVERPOOL}
\DpName{U.Amaldi}{MILANO2}
% ...
\normalsize
\endgroup
\titlefoot{Department of Physics and Astronomy, Iowa State
     University, Ames IA 50011-3160, USA
    \label{AMES}}
\titlefoot{Physics Department, Universiteit Antwerpen,
     Universiteitsplein 1, B-2610 Antwerpen, Belgium \\
     \indent~~and IIHE, ULB-VUB,
     Pleinlaan 2, B-1050 Brussels, Belgium \\
     \indent~~and Facult\'e des Sciences,
     Univ. de l'Etat Mons, Av. Maistriau 19, B-7000 Mons, Belgium
    \label{AIM}}
\titlefoot{Physics Laboratory, University of Athens, Solonos Str.
     104, GR-10680 Athens, Greece
    \label{ATHENS}}
\titlefoot{Department of Physics, University of Bergen,
     All\'egaten 55, NO-5007 Bergen, Norway
    \label{BERGEN}}
\titlefoot{Dipartimento di Fisica, Universit\`a di Bologna and INFN,
     Via Irnerio 46, IT-40126 Bologna, Italy
    \label{BOLOGNA}}
\titlefoot{Centro Brasileiro de Pesquisas F\'{\i}sicas, rua Xavier Sigaud 150,
     BR-22290 Rio de Janeiro, Brazil \\
     \indent~~and Depto. de F\'{\i}sica, Pont. Univ. Cat\'olica,
     C.P. 38071 BR-22453 Rio de Janeiro, Brazil \\
     \indent~~and Inst. de F\'{\i}sica, Univ. Estadual do Rio de Janeiro,
     rua S\~{a}o Francisco Xavier 524, Rio de Janeiro, Brazil
    \label{BRASIL}}
\titlefoot{Coll\`ege de France, Lab. de Physique Corpusculaire, IN2P3-CNRS,
     FR-75231 Paris Cedex 05, France
    \label{CDF}}
\titlefoot{CERN, CH-1211 Geneva 23, Switzerland
    \label{CERN}}
\titlefoot{Institut de Recherches Subatomiques, IN2P3 - CNRS/ULP - BP20,
     FR-67037 Strasbourg Cedex, France
    \label{CRN}}
\titlefoot{Now at DESY-Zeuthen, Platanenallee 6, D-15735 Zeuthen, Germany
    \label{DESY}}
\titlefoot{Institute of Nuclear Physics, N.C.S.R. Demokritos,
     P.O. Box 60228, GR-15310 Athens, Greece
    \label{DEMOKRITOS}}
\titlefoot{FZU, Inst. of Phys. of the C.A.S. High Energy Physics Division,
     Na Slovance 2, CZ-180 40, Praha 8, Czech Republic
    \label{FZU}}
\titlefoot{Dipartimento di Fisica, Universit\`a di Genova and INFN,
     Via Dodecaneso 33, IT-16146 Genova, Italy
    \label{GENOVA}}
\titlefoot{Institut des Sciences Nucl\'eaires, IN2P3-CNRS, Universit\'e
     de Grenoble 1, FR-38026 Grenoble Cedex, France
    \label{GRENOBLE}}
\titlefoot{Helsinki Institute of Physics and Department of Physical Sciences,
     P.O. Box 64, FIN-00014 University of Helsinki, 
     \indent~~Finland
    \label{HELSINKI}}
\titlefoot{Joint Institute for Nuclear Research, Dubna, Head Post
     Office, P.O. Box 79, RU-101 000 Moscow, Russian Federation
    \label{JINR}}
\titlefoot{Institut f\"ur Experimentelle Kernphysik,
     Universit\"at Karlsruhe, Postfach 6980, DE-76128 Karlsruhe,
     Germany
    \label{KARLSRUHE}}
\titlefoot{Institute of Nuclear Physics PAN,Ul. Radzikowskiego 152,
     PL-31142 Krakow, Poland
    \label{KRAKOW1}}
\titlefoot{Faculty of Physics and Nuclear Techniques, University of Mining
     and Metallurgy, PL-30055 Krakow, Poland
    \label{KRAKOW2}}
\titlefoot{Universit\'e de Paris-Sud, Lab. de l'Acc\'el\'erateur
     Lin\'eaire, IN2P3-CNRS, B\^{a}t. 200, FR-91405 Orsay Cedex, France
    \label{LAL}}
\titlefoot{School of Physics and Chemistry, University of Lancaster,
     Lancaster LA1 4YB, UK
    \label{LANCASTER}}
\titlefoot{LIP, IST, FCUL - Av. Elias Garcia, 14-$1^{o}$,
     PT-1000 Lisboa Codex, Portugal
    \label{LIP}}
\titlefoot{Department of Physics, University of Liverpool, P.O.
     Box 147, Liverpool L69 3BX, UK
    \label{LIVERPOOL}}
\titlefoot{Dept. of Physics and Astronomy, Kelvin Building,
     University of Glasgow, Glasgow G12 8QQ
    \label{GLASGOW}}
\titlefoot{LPNHE, IN2P3-CNRS, Univ.~Paris VI et VII, Tour 33 (RdC),
     4 place Jussieu, FR-75252 Paris Cedex 05, France
    \label{LPNHE}}
\titlefoot{Department of Physics, University of Lund,
     S\"olvegatan 14, SE-223 63 Lund, Sweden
    \label{LUND}}
\titlefoot{Universit\'e Claude Bernard de Lyon, IPNL, IN2P3-CNRS,
     FR-69622 Villeurbanne Cedex, France
    \label{LYON}}
\titlefoot{Dipartimento di Fisica, Universit\`a di Milano and INFN-MILANO,
     Via Celoria 16, IT-20133 Milan, Italy
    \label{MILANO}}
\titlefoot{Dipartimento di Fisica, Univ. di Milano-Bicocca and
     INFN-MILANO, Piazza della Scienza 2, IT-20126 Milan, Italy
    \label{MILANO2}}
\titlefoot{IPNP of MFF, Charles Univ., Areal MFF,
     V Holesovickach 2, CZ-180 00, Praha 8, Czech Republic
    \label{NC}}
\titlefoot{NIKHEF, Postbus 41882, NL-1009 DB
     Amsterdam, The Netherlands
    \label{NIKHEF}}
\titlefoot{National Technical University, Physics Department,
     Zografou Campus, GR-15773 Athens, Greece
    \label{NTU-ATHENS}}
\titlefoot{Physics Department, University of Oslo, Blindern,
     NO-0316 Oslo, Norway
    \label{OSLO}}
\titlefoot{Dpto. Fisica, Univ. Oviedo, Avda. Calvo Sotelo
     s/n, ES-33007 Oviedo, Spain
    \label{OVIEDO}}
\titlefoot{Department of Physics, University of Oxford,
     Keble Road, Oxford OX1 3RH, UK
    \label{OXFORD}}
\titlefoot{Dipartimento di Fisica, Universit\`a di Padova and
     INFN, Via Marzolo 8, IT-35131 Padua, Italy
    \label{PADOVA}}
\titlefoot{Rutherford Appleton Laboratory, Chilton, Didcot
     OX11 OQX, UK
    \label{RAL}}
\titlefoot{Dipartimento di Fisica, Universit\`a di Roma II and
     INFN, Tor Vergata, IT-00173 Rome, Italy
    \label{ROMA2}}
\titlefoot{Dipartimento di Fisica, Universit\`a di Roma III and
     INFN, Via della Vasca Navale 84, IT-00146 Rome, Italy
    \label{ROMA3}}
\titlefoot{DAPNIA/Service de Physique des Particules,
     CEA-Saclay, FR-91191 Gif-sur-Yvette Cedex, France
    \label{SACLAY}}
\titlefoot{Instituto de Fisica de Cantabria (CSIC-UC), Avda.
     los Castros s/n, ES-39006 Santander, Spain
    \label{SANTANDER}}
\titlefoot{Inst. for High Energy Physics, Serpukov
     P.O. Box 35, Protvino, (Moscow Region), Russian Federation
    \label{SERPUKHOV}}
\titlefoot{J. Stefan Institute, Jamova 39, SI-1000 Ljubljana, Slovenia
     and Laboratory for Astroparticle Physics,\\
     \indent~~Nova Gorica Polytechnic, Kostanjeviska 16a, SI-5000 Nova Gorica, Slovenia, \\
     \indent~~and Department of Physics, University of Ljubljana,
     SI-1000 Ljubljana, Slovenia
    \label{SLOVENIJA}}
\titlefoot{Fysikum, Stockholm University,
     Box 6730, SE-113 85 Stockholm, Sweden
    \label{STOCKHOLM}}
\titlefoot{Dipartimento di Fisica Sperimentale, Universit\`a di
     Torino and INFN, Via P. Giuria 1, IT-10125 Turin, Italy
    \label{TORINO}}
\titlefoot{INFN,Sezione di Torino and Dipartimento di Fisica Teorica,
     Universit\`a di Torino, Via Giuria 1,
     IT-10125 Turin, Italy
    \label{TORINOTH}}
\titlefoot{Dipartimento di Fisica, Universit\`a di Trieste and
     INFN, Via A. Valerio 2, IT-34127 Trieste, Italy \\
     \indent~~and Istituto di Fisica, Universit\`a di Udine,
     IT-33100 Udine, Italy
    \label{TU}}
\titlefoot{Univ. Federal do Rio de Janeiro, C.P. 68528
     Cidade Univ., Ilha do Fund\~ao
     BR-21945-970 Rio de Janeiro, Brazil
    \label{UFRJ}}
\titlefoot{Department of Radiation Sciences, University of
     Uppsala, P.O. Box 535, SE-751 21 Uppsala, Sweden
    \label{UPPSALA}}
\titlefoot{IFIC, Valencia-CSIC, and D.F.A.M.N., U. de Valencia,
     Avda. Dr. Moliner 50, ES-46100 Burjassot (Valencia), Spain
    \label{VALENCIA}}
\titlefoot{Institut f\"ur Hochenergiephysik, \"Osterr. Akad.
     d. Wissensch., Nikolsdorfergasse 18, AT-1050 Vienna, Austria
    \label{VIENNA}}
\titlefoot{Inst. Nuclear Studies and University of Warsaw, Ul.
     Hoza 69, PL-00681 Warsaw, Poland
    \label{WARSZAWA}}
\titlefoot{Fachbereich Physik, University of Wuppertal, Postfach
     100 127, DE-42097 Wuppertal, Germany \\
\noindent
{$^\dagger$~deceased}
    \label{WUPPERTAL}}
\addtolength{\textheight}{-10mm}
\addtolength{\footskip}{5mm}
\clearpage
\headsep 30.0pt
\end{titlepage}
%
\pagenumbering{arabic} 
\setcounter{footnote}{0} %
\large
\section{Introduction}

\begin{figure}[ht!]
\begin{center}
\vspace*{1.5cm}
\hspace*{-1.cm}
\epsfig{file=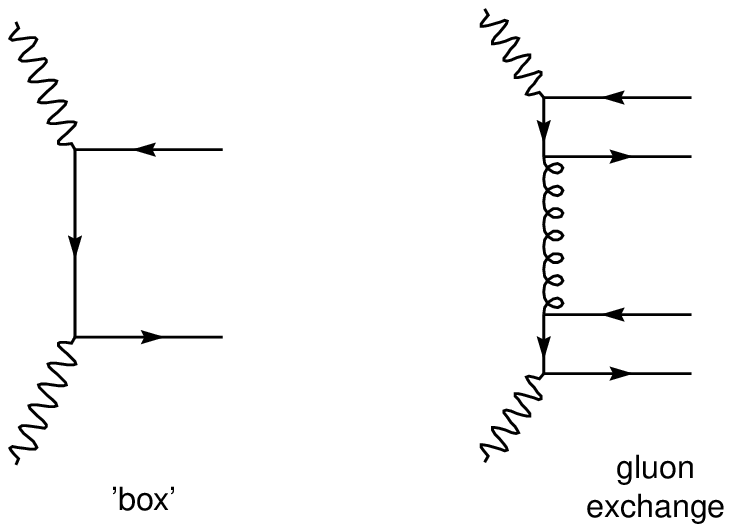,width=11.0cm}
\end{center}
\vspace*{-2.5cm}
\caption[]{Main diagrams corresponding to the \gghadt process.}
\label{fig:fig1}
\end{figure}

This paper presents the study of double-tagged two-photon 
interactions $e^+e^-\rightarrow e^+e^-\gamma^*\gamma^* \rightarrow 
e^+e^-X$ (where X is either a muon pair or hadrons) with the DELPHI
detector \cite{delphi} at the CERN LEPII collider.
Both scattered electrons\footnote{Throughout this paper, electron stands both for 
electron and positron. Asterisk over $\gamma$ symbol explicitly indicates
that the photon is highly virtual.}
are detected by the Small angle TIle Calorimeter (STIC).
Compared to the untagged or single-tagged modes, with both or one of 
the electrons escaping detection, this mode of gamma-gamma collision 
has the advantage that the kinematics of the interaction is well 
defined by the measurement of the energies and scattering angles of the tagged particles. 
The production of muon pairs is described by QED. 
Similarly, multihadron production is expected to be described by QPM, but only
in a first approximation. If the virtualities of the photons are large enough, 
it is predicted that there should be a large contribution from processes
with (multi)gluon exchange between $q\bar{q}$ dipoles \cite{theor},
which is described by the BFKL equation \cite{bfkl}. 
Two-photon interactions are therefore a suitable process to investigate 
BFKL dynamics.
Figure \ref{fig:fig1} shows the main diagrams relevant to the analysis.

\par
\begin{figure}[ht!]
\begin{center}
\hspace*{-1.cm}
\epsfig{file=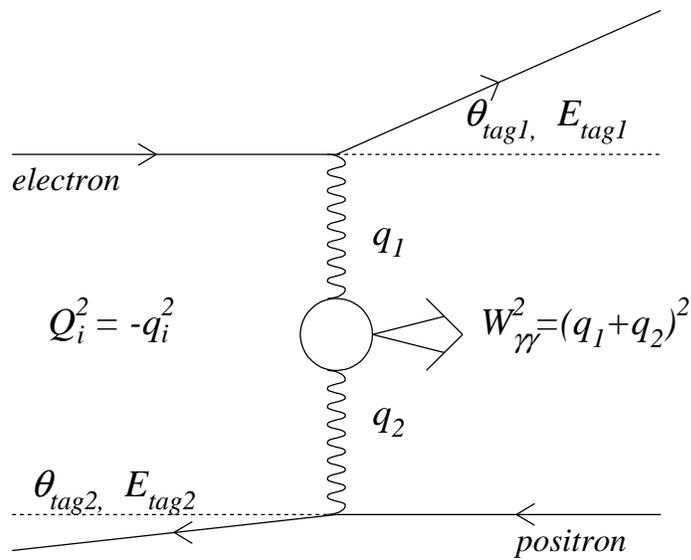,width=10.cm}
\end{center}
\caption[]{The kinematics of \agg interactions.}
\label{fig:fig2}
\end{figure}

 
The kinematics of the process is illustrated in Figure \ref{fig:fig2}.
We use the following notations: $p_i$ (i=1,2) are the four-momenta of 
the beam electrons, $\sqrt{s}$ is the $e^+e^-$ centre-of-mass energy,  
$E_{beam}$ is the beam energy; the four-momenta of the scattered electrons,
their polar angles and their energies are $p_i^{'}$, $\theta_i$ and $E_i$ 
respectively.

\par The variables relevant to this study are the virtualities of the photons, 
$Q^2_i$, the invariant mass of the two photons $\wgg$ and a dimensionless 
variable $Y$:
\begin{itemize}
\item $Q^2_i = -q^2_i = -(p_i - p_i^{'})^2 = 4E_iE_{beam}\sin^2(\theta_i/2)$;
\item $\wgg^2 = -(q_1 + q_2)^2 \simeq sy_1y_2$ with
$y_i=1-(E_i/E_{beam})\cos^2(\theta_i/2)$;
\item $Y=\ln(\wgg^2/\sqrt{Q^2_1Q^2_2})$.
\end{itemize}
\par The $Y$ variable is used to compare the multihadron data with the BFKL predicted
cross-section with the conditions $\wgg^2 \gg Q^2_i$ and
$\mid\ln(Q^2_1/Q^2_2)\mid < 1$, 
where the second condition is needed to select virtualities of the photons
of the same order.
\par The analysis is divided into two parts: the study of the production of muon pairs 
aims at comparing the data with the well-known QED model and at tuning
the experimental cuts, while the multihadron production is used 
to measure the cross-section \sgg and to
compare it with the BFKL predictions. The models used for each part and the
background estimations are described separately. \\
\section{Detector and data sample}
\par A detailed description of the DELPHI detector and of its performance 
is presented in ref. \cite{delphi}: here only the components relevant 
to the present analysis will be briefly mentioned. 
\par The scattered electrons are detected in the luminosity monitor STIC, 
which covers the region from 29 mrad to 185 mrad in the polar angle 
$\theta$ \footnote{The origin of the DELPHI reference system was at the centre of the detector.
It coincides with the ideal interaction point. The z-axis was parallel to the $e^-$ beam,
the x-axis pointed horizontally to the centre of the LEP ring and the y-axis
was vertically upward. The co-ordinates $R,\phi$,z formed a cylindrical coordinate system
and $\theta$ was the polar angle with respect to the z-axis.},
with $R\phi$ segmentation of 3 cm $\times 22.5^\circ$ \cite{stic}.
Given the energy and angular resolution of the STIC calorimeter,
the $Q^2$ resolution varies between 1 GeV$^2$ and 2.5 GeV$^2$ in the 
$Q^2$ domain of the present analysis ($Q^2_i$ between 10 GeV$^2$ and 200 GeV$^2$).
\par Charged particles are detected in the barrel tracking system 
comprising the Silicon Tracker (ST), 
the Inner Detector (ID), the Time Projection Chamber (TPC), and the Outer Detector. 
In the endcap regions, they are detected by the ST, by the TPC down to 20$^\circ$
in polar angle, by the ID down to 15$^\circ$ and 
by the Forward Chambers A and B. All detectors are located inside 
a superconducting solenoid providing a uniform magnetic field 
of 1.23 T parallel to the axis of the colliding $e^+e^-$ beams.
The combined momentum resolution provided by the tracking
system is a few per-mille in the momentum range of this study.
\par Muon tagging is performed with the Barrel and the Forward 
muon drift chambers, and with the Surround Muon Chambers based on 
limited streamer tubes, which cover the gap between the previous two.
\par The study is done with the DELPHI data  
taken during 1998-2000 at $e^+e^-$ centre-of-mass energies 
from 189 GeV to 209 GeV, corresponding to an integrated 
luminosity of 550 pb$^{-1}$, with the subdetectors relevant 
for the analysis all fully operational.
\par Simulated events for the physics processes and backgrounds are 
generated at the different centre-of-mass energies and passed through the 
full DELPHI simulation and reconstruction chain.

\section{Study of \boldmath{\ggeemm} interactions}

\subsection{Data analysis}
\par The following criteria are used to select \ggmmdt events:
\begin{itemize}
\item There are two clusters with energy deposition $E_i$ greater than 
30 GeV, one in each arm of the STIC
\footnote{The choice of the cutoff for 
the minimum energy of the tagged particles has to be
made carefully. It should be as small as possible since low electron energy 
corresponds to large values of $\wgg$ and $Y$ (which are important in the 
multihadron case, see below). At the same time the lowering of this cut 
leads to an increase of the off-momentum background and thus decreases 
the accuracy of the measurement.}, and the polar angle $\theta_i$
(or (180$^\circ$-$\theta_i$) if the angle is more than 90$^\circ$) exceeds 
$2.2^\circ$ for each cluster;
\item $Q^2_i$ is   
between 10 GeV$^2$ and 200 GeV$^2$ for both tagged particles;
\item The acollinearity of the scattered electrons is
above 0.2 degrees. This criterion removes
a superposition between Bhabha events and untagged \ggmmnt events;
\item Each event contains two charged particles with zero net charge
and invariant mass between 2 GeV/${\it c}^2$ and 50 GeV/${\it c}^2$.
Particles are considered if their momentum is greater than 400 MeV/${\it c}$,
their polar angle is within the interval $20^\circ$ - $160^\circ$ and
their impact parameters are smaller than 4~ cm in $R\phi$ 
and 10~ cm in $z$;
\item At least one of the charged particles is identified as a `standard` or
`tight` muon by the DELPHI tagging algorithm \cite{delphi}.
\end{itemize}

\par The number of selected \ggeemm events is 226.
Double-tagged events are triggered
either by the STIC trigger component or by a single charged particle track 
component \cite{trig}.
The trigger efficiency has been calculated using the redundancy of 
the trigger together with independent calculations based on the parameterization 
of the single track efficiency \cite{trig}, and is found to be larger than 99\%.  
\par The BDKRC event generator 
\cite{bdkrc}, including the full set of QED diagrams and all fermion masses,
is used for the Monte Carlo simulation of \ggmmdt events.
The processes corresponding to diagrams other than the diagram of two-photon interactions
(multiperipheral) are found to give a contribution of about 2 percent.
The number of multiperipheral events is expected to be 194.
\par The following sources of background are considered:
\begin{itemize}
\item The coincidence of a gamma-gamma event with an off-momentum electron.
The probability of such coincidences, averaged over the data from different years,  
is calculated with \ggmmnt events to be (0.0016 $\pm 0.0002)$.
Using this value, the background from
a superimposition of two off-momentum electrons with an untagged \ggmmnt event 
turns out to be negligible;
\item The coincidence 
of one off-momentum electron with a \ggmmst single-tagged event, i.e. when one scattered
electron is detected in the STIC while the other one is an off-momentum electron. 
Usually this background is evaluated by convoluting the \ggmmst Monte Carlo 
simulation with the spectrum of off-momentum electrons. This requires the 
appropriate description of the single-tagged data by the simulation and thus 
it is model-dependent. The following approach avoids the problem and 
calculates the background directly from the data.
The cuts as listed above are applied with one difference:
events with two electrons in the same STIC arm and none in the other are 
selected. They include one off-momentum electron. Then 
one of the electrons is rotated to the ``empty'' STIC arm, i.e. its $p_z$ component is 
inverted. The background coming from the coincidence of one off-momentum with 
a \ggmmst single-tagged event is thus estimated to be $(15\pm4)$ events;
\item The background from \ggttdt events is estimated as $(23\pm2)$ events
by using the TWOGAM event generator \cite{twogam}.
\end{itemize}
\par The overall background is thus estimated as $(38\pm4)$ events.

\par
\begin{figure}[ht!]
\begin{tabular}{cc}
\begin{minipage}[h]{9cm}
\vspace*{-2cm}
\hspace*{-2cm}
\mbox{\epsfig{figure=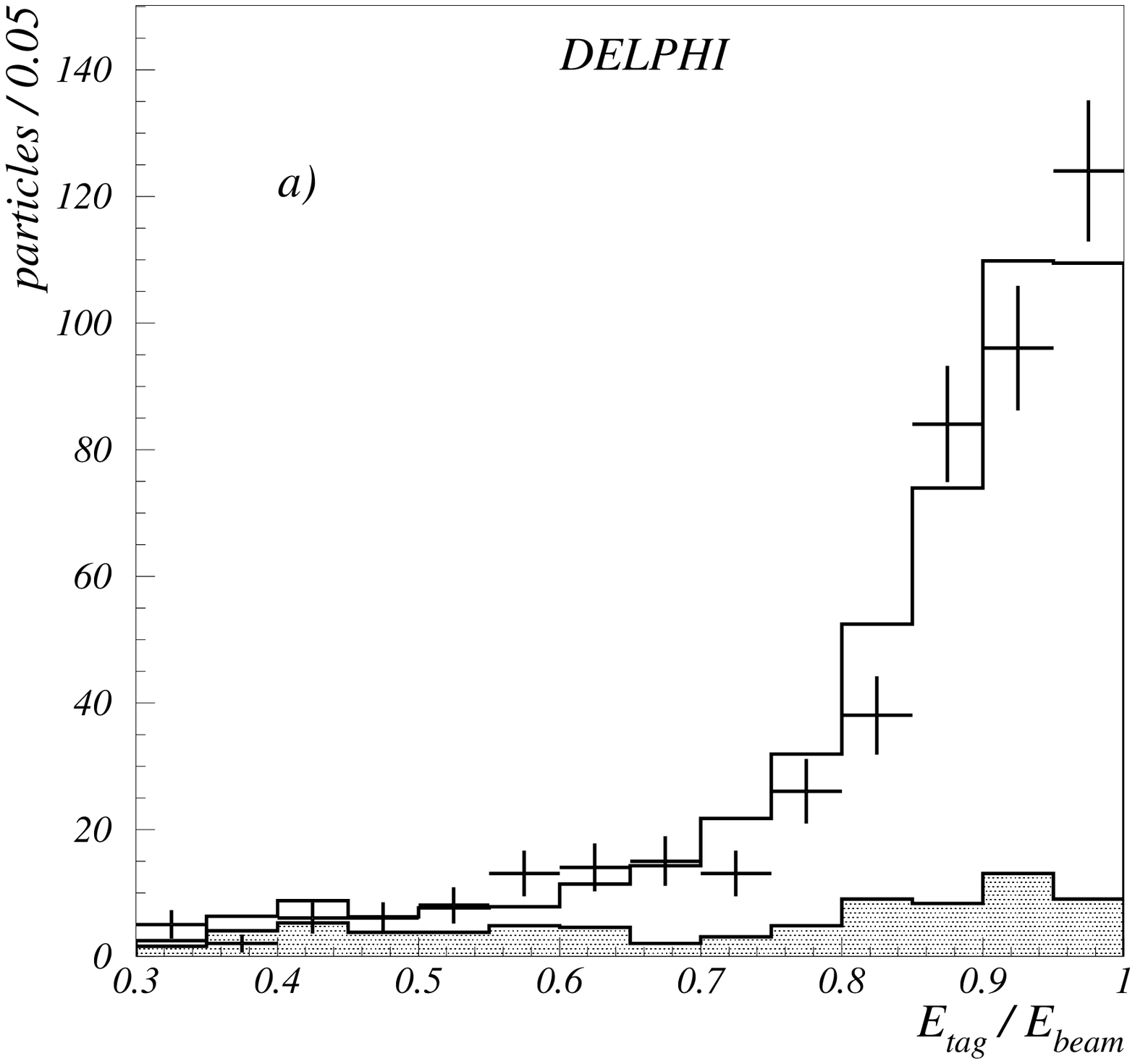,width=8.5cm}}
\end{minipage}
\begin{minipage}[h]{9cm}
\vspace*{-2cm}
\hspace*{-2.5cm}
\mbox{\epsfig{figure=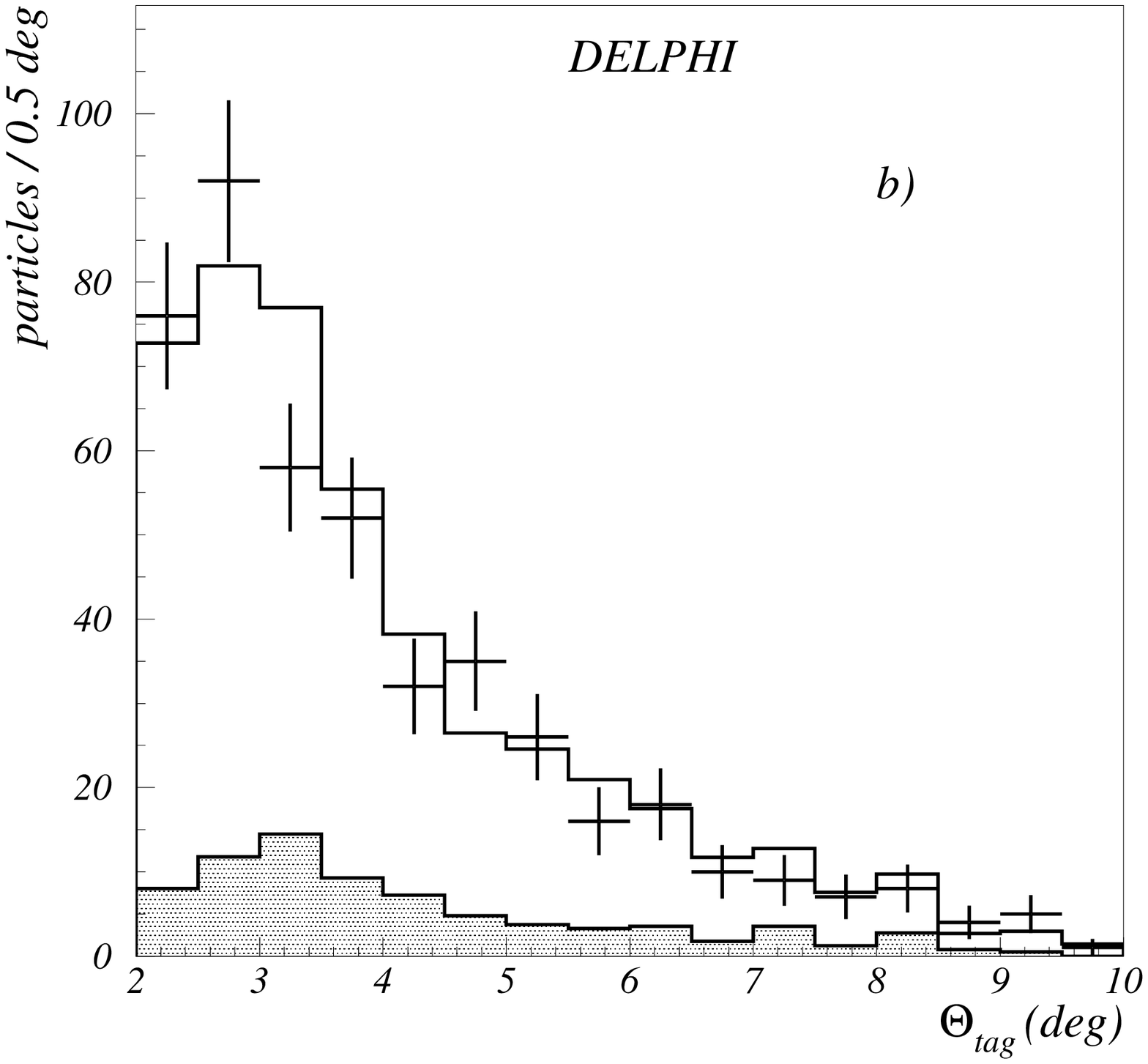,width=8.5cm}}
\end{minipage}
\end{tabular} \\ \\ \\

\begin{tabular}{cc}
\begin{minipage}[h]{9cm}
\vspace*{1cm}
\hspace*{-2.cm}
\mbox{\epsfig{figure=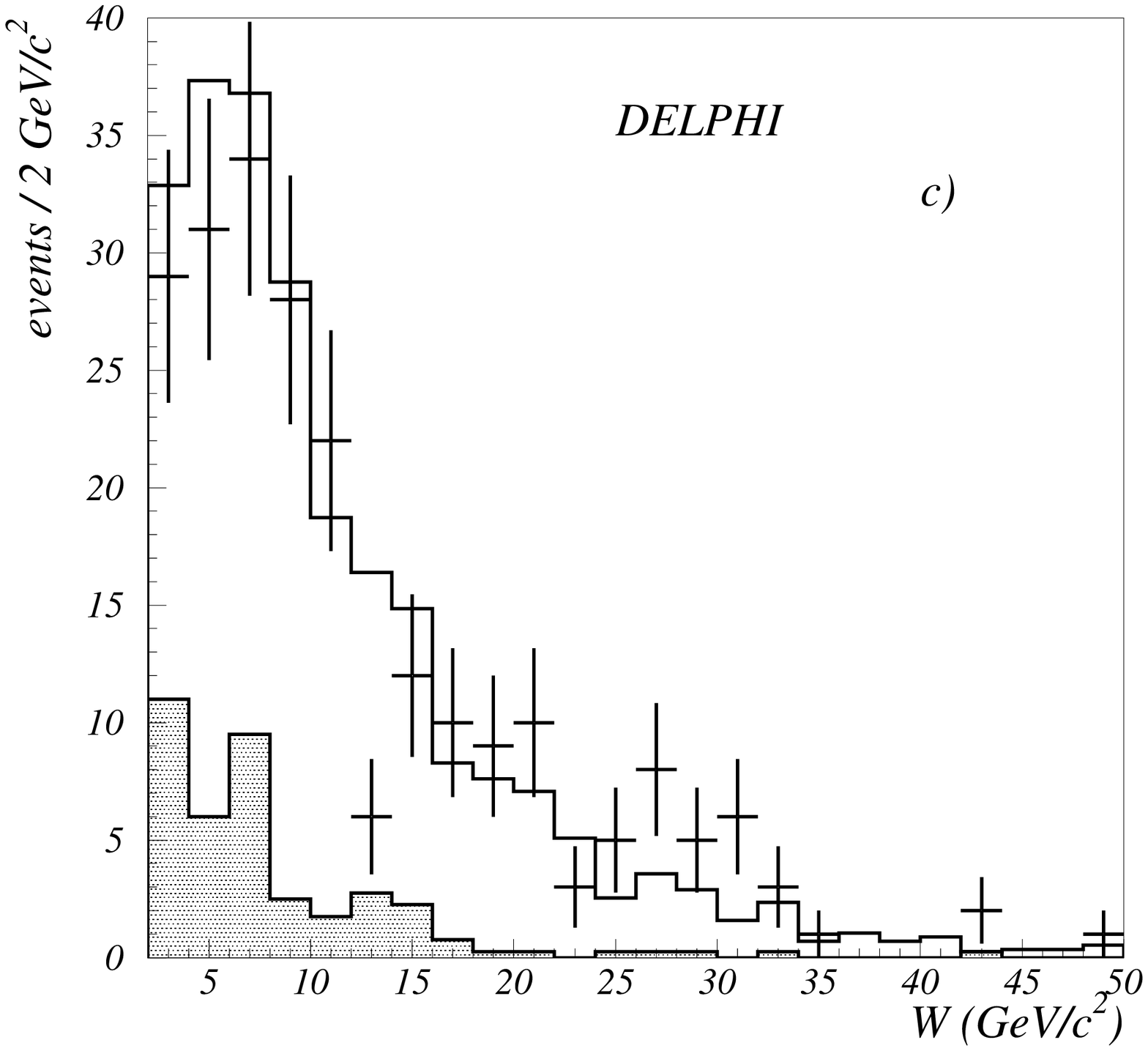,width=8.5cm}}
\end{minipage}
\begin{minipage}[h]{9cm}
\vspace*{1cm}
\hspace*{-2.5cm}
\mbox{\epsfig{figure=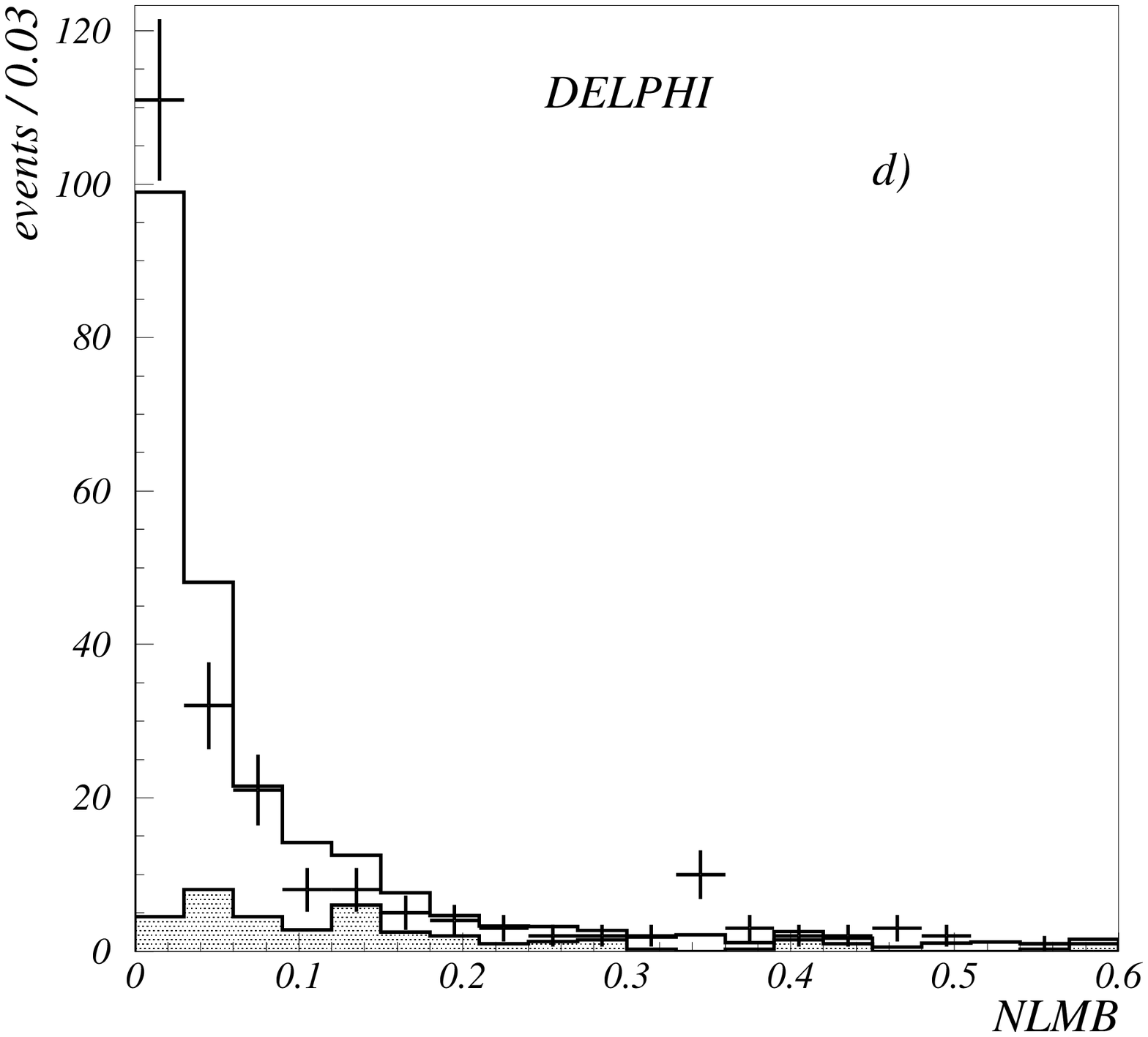,width=8.5cm}}
\end{minipage}
\end{tabular}
\caption[]{(a) Normalized tagged particle energy $E_i/E_{beam}$,
(b) tagged particle polar angle $\theta_i$ (two entries per
event for both), (c) invariant mass 
of the muon pair $W_{\mu\mu}$ and (d) $NLMB$ variable.
The data are shown with error bars; the histograms are the sum of the 
BDKRC simulated events and of the estimated background. 
The hatched areas represent the background contamination.}
\label{fig:fig3}
\end{figure}

 Figure \ref{fig:fig3} shows the distributions of the tagged particle 
energies\footnote{If the measured value of the electron energy is greater than
$E_{beam}$, it is changed to ($E_{beam}$-0.5~ GeV) to be able to
calculate the $\gamma^*\gamma^*$ invariant mass.} normalized to the beam energy,
their polar angles (two entries per event for both histograms), the 
invariant mass of the muon pair, $W_{\mu\mu}$, calculated from the muon 
4-momenta, and the distribution of the normalized 
longitudinal momentum balance defined as \\
\hspace*{1cm} $NLMB = \mid p_{z,tag1}+p_{z,tag2}+p_{z,X} \mid / E_{beam}$, \\
where $p_{z,X}$ is the $z-$component of the momentum of the system produced 
in the \agg collision. This variable has to be peaked at zero for  
well-reconstructed events and it is sensitive to the final state radiation.

\par
\begin{figure}[ht!]
\begin{tabular}{cc}
\begin{minipage}[h]{9cm}
\hspace*{-2.cm}
\mbox{\epsfig{figure=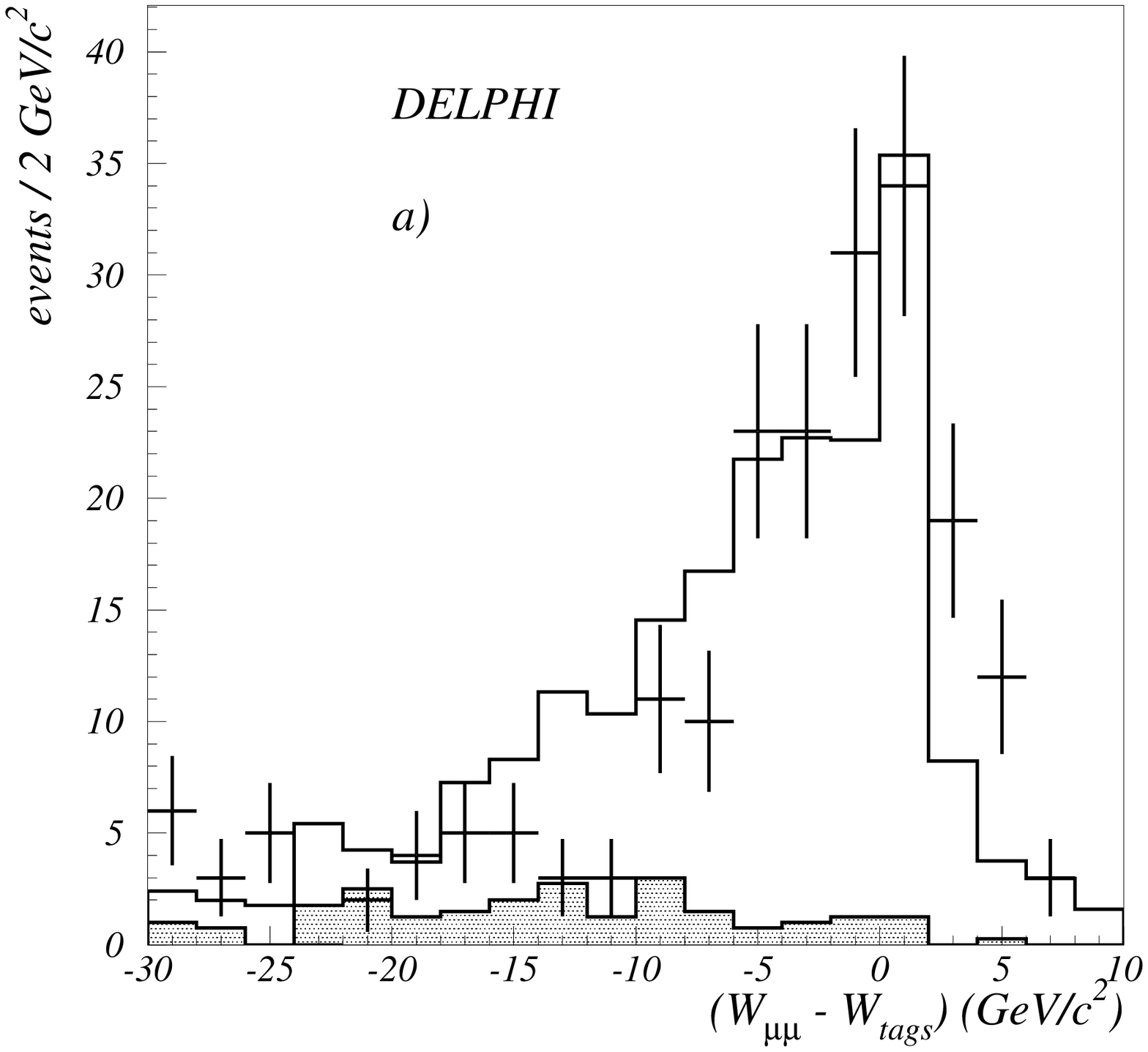,width=8.5cm}}
\end{minipage}
\begin{minipage}[h]{9cm}
\hspace*{-2.5cm}
\mbox{\epsfig{figure=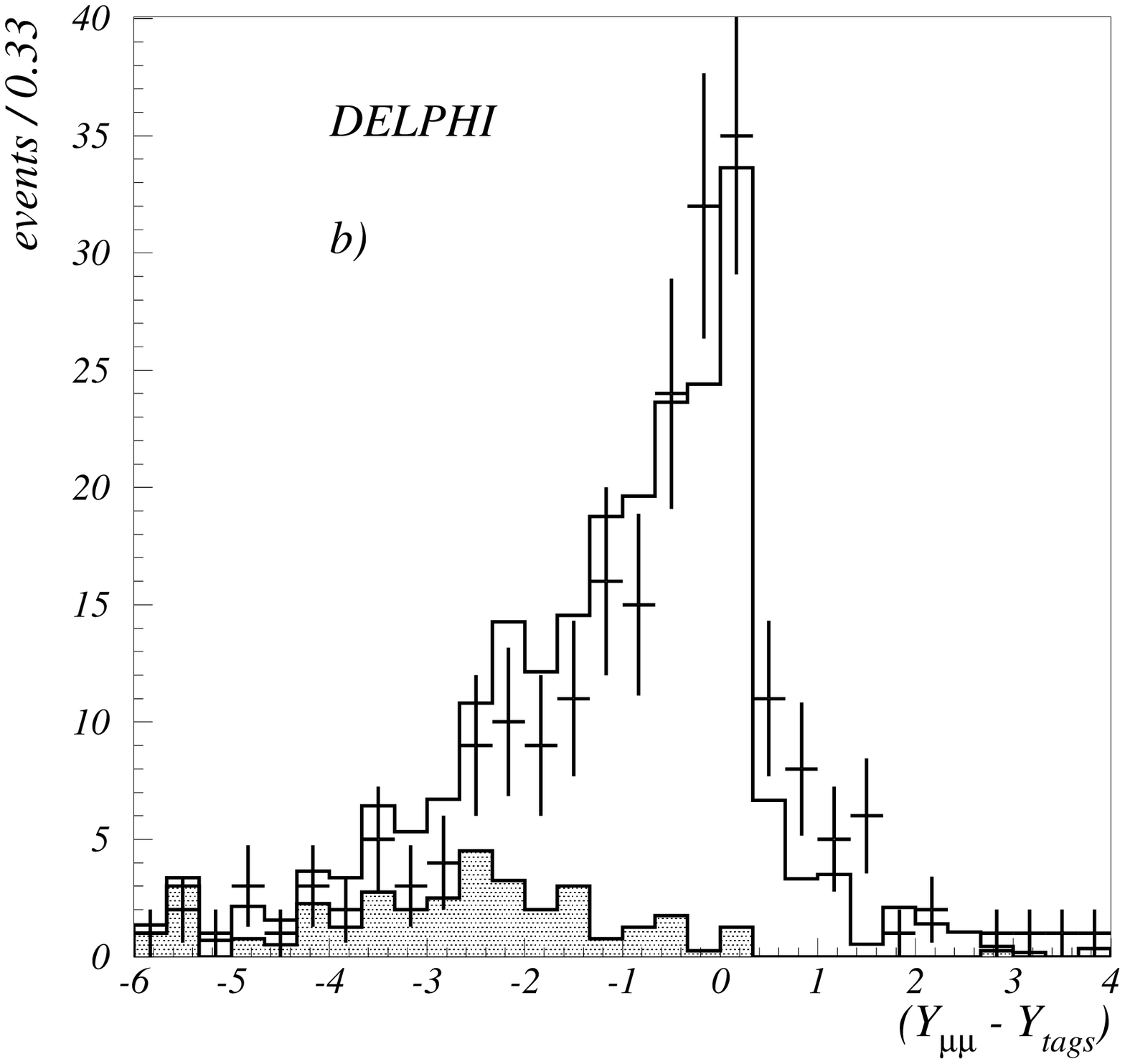,width=8.5cm}}
\end{minipage}
\end{tabular}
\caption[]{(a) Difference between the invariant mass of the muon pair, 
$W_{\mu\mu}$, and the invariant mass calculated from the tagged particles.
(b) The same for the $Y$ variable.
The data are shown with error bars; the dashed histograms are the sum of the 
BDKRC simulated events and of the estimated background.
The hatched areas represent the background contamination.}
\label{fig:fig4}
\end{figure}

 The analysis of \ggmmdt events permits the quality of the
reconstruction to be examined. The kinematics of the gamma-gamma system is completely 
determined by the measurements of the muons and of the tagged  electrons in
this exclusive channel, and some quantities can be calculated either from 
the tagged particles or from the muons. 
Figure \ref{fig:fig4}a (b) shows the difference between the gamma-gamma invariant 
mass ($Y$ variable) calculated from the muon 4-momenta $W_{\mu\mu} 
(Y_{\mu\mu}$) and that reconstructed from the tagged particles' measurements 
$\wgg (Y_{\gamma\gamma})$. The asymmetry of the distributions is due to
radiative corrections, as it has been verified by Monte Carlo simulation.
These comparisons show that the use of tagged particles is a good approximation to
calculate the kinematic variables of the gamma-gamma system.

\subsection{Results}
\par The selected data sample is used for the measurement of the
cross-section (\see\hspace*{-1mm}) of the reaction \ggeemm. The corrections
for the detector acceptance and efficiency are done with
the BDKRC simulated events. The statistical uncertainty in the MC simulation is
included in the systematic error. Additional systematic uncertainties are evaluated by
varying the selection criteria on the tagging particles. Systematic uncertainty coming 
from the muon identification procedure is negligible (second item of \cite{delphi},
p.96 and references therein).
\par The measured total \see
cross-section is $(1.38\pm0.12(stat)\pm0.06(syst))$ pb for virtualities 
of the interacting photons, $Q^2_i$, between 10 GeV$^2$ and 200 GeV$^2$
and for invariant mass $W_{\mu\mu}$ between 2 GeV/${\it c}^2$ and 50 GeV/${\it c}^2$.
The QED expectation, including radiative corrections, is $(1.36\pm0.01)$ pb. 
The cross-section calculated without radiative corrections is about 8\% lower.
\par
\begin{figure}[ht!]
\begin{center}
\mbox{\epsfig{figure=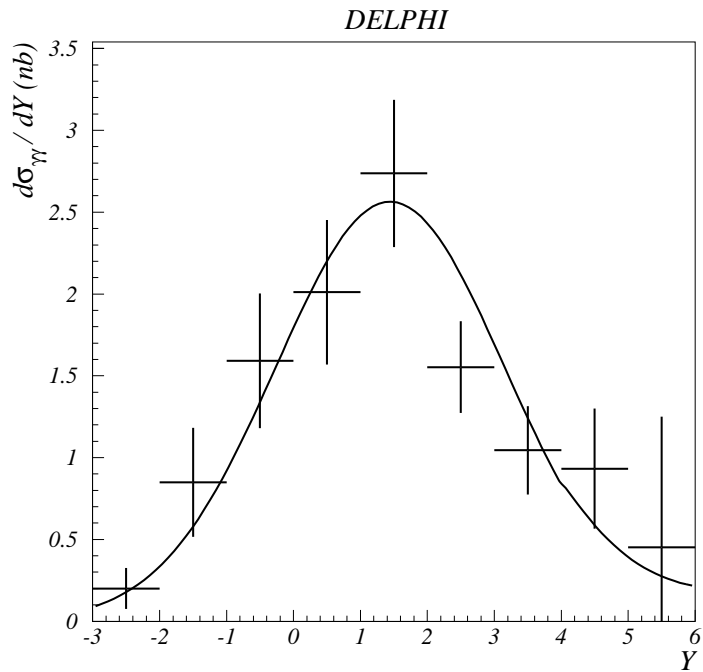,width=10cm}}
\end{center}
\caption[]{The differential cross-section for the reaction \ggmmdt. 
The data are shown with error bars. The solid line shows the QED expectation.}
\label{fig:fig5}
\end{figure}

 The \see cross-section
can be expressed via the flux of photons with different 
polarization and the corresponding partial cross-sections of the \ggmmdt 
interaction. This extraction procedure is described in the next section 
for multihadron production. Here only the result for the
differential \ggmmdt cross-section as a function of $Y$
is shown in figure \ref{fig:fig5}. 
There is good agreement between the measurements and the QED predictions. 

\section{Study of \boldmath{\ggeeha} interactions}

\subsection{Data analysis}
\par 
The selection criteria for tagging electrons and for the charged particles
are the same as described in the previous section.  
The sample of \gghadt events is then selected by the following criterion:
\begin{itemize}
\item Each event contains at least 3 charged particles with the invariant 
mass calculated from the particles' 4-momenta, $W_{had}$, larger than 2 GeV/${\it c}^2$;
\end{itemize}
The following additional cuts are applied to suppress background events:
\begin{itemize}
\item If the energy of one cluster in STIC, normalized to the beam energy, is
larger than 0.85 then the energy of another cluster has to be below 0.5. 
This cut is intended to suppress the contamination coming from \eehadt events;
\item The thrust value of the charged particles, calculated in their
centre-of-mass system, is less than 0.98 for the events with charged 
multiplicity below 5. The cut removes most of the \ggttdt events.
\end{itemize}

\par After these requirements, 434 events have been selected. 
Again the trigger efficiency \cite{trig} can be estimated from the redundancy 
of the trigger and from a parameterization of the single track efficiency, 
and is larger than 99\%.
\par The event generators used to simulate the \agg events and the background 
processes are listed below as well as the respective expected contributions:
\begin{itemize}
\item TWOGAM (version 2.02) \cite{twogam} and PYTHIA (version 6.205) 
\cite{pyt} event generators (both include radiative corrections) 
are used to simulate \agg interactions.
The expectations are $(331\pm8)$ and $(330\pm8)$ events, respectively.
The Monte Carlo generators include the quark-parton model (QPM) part
and also the leading-order predictions for the resolved photon contribution;
\item The background coming from the process \eehadt is simulated with the
{\mbox KK2f} generator (version 4.14) \cite{kk2f} and its contribution 
is estimated to be ($27\pm3$) events;
\item The background of $\tau$ pairs produced 
in $e^+e^-$ annihilation is found to be negligible;
\item The contamination of $\tau$ pairs produced in the two-photon interactions
is evaluated as ($26\pm3$) events by using the TWOGAM program;
\item The coincidence of an off-momentum electron with a \gghast single-tagged 
event is evaluated as ($5\pm2$) events by using the same approach as described 
in the previous section.
\end{itemize}
\par
\begin{figure}[ht!]
\begin{tabular}{cc}
\begin{minipage}[h]{9cm}
\vspace*{-2cm}
\hspace*{-2cm}
\mbox{\epsfig{figure=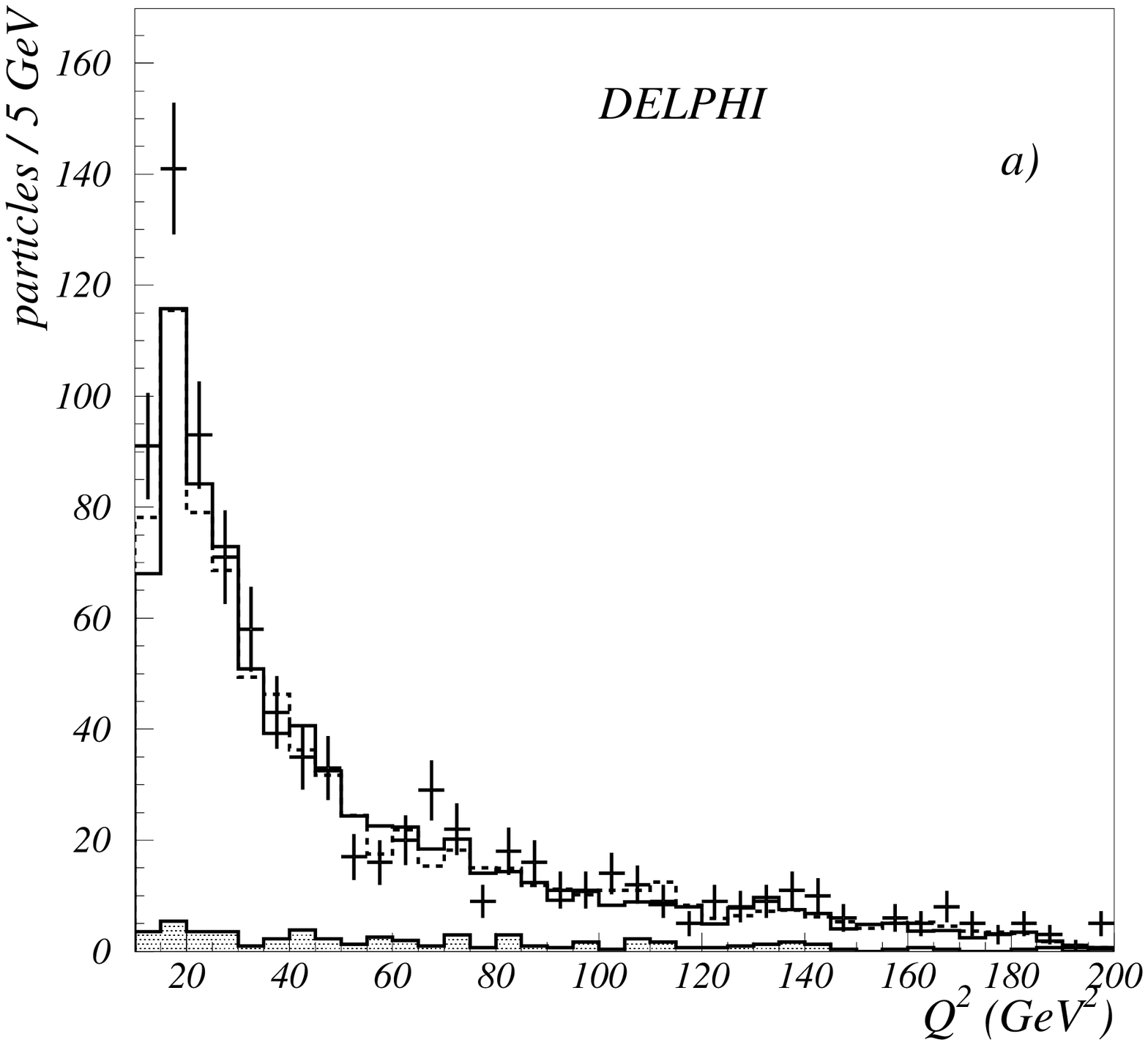,width=8.5cm}}
\end{minipage}
\begin{minipage}[h]{9cm}
\vspace*{-2cm}
\hspace*{-2.5cm}
\mbox{\epsfig{figure=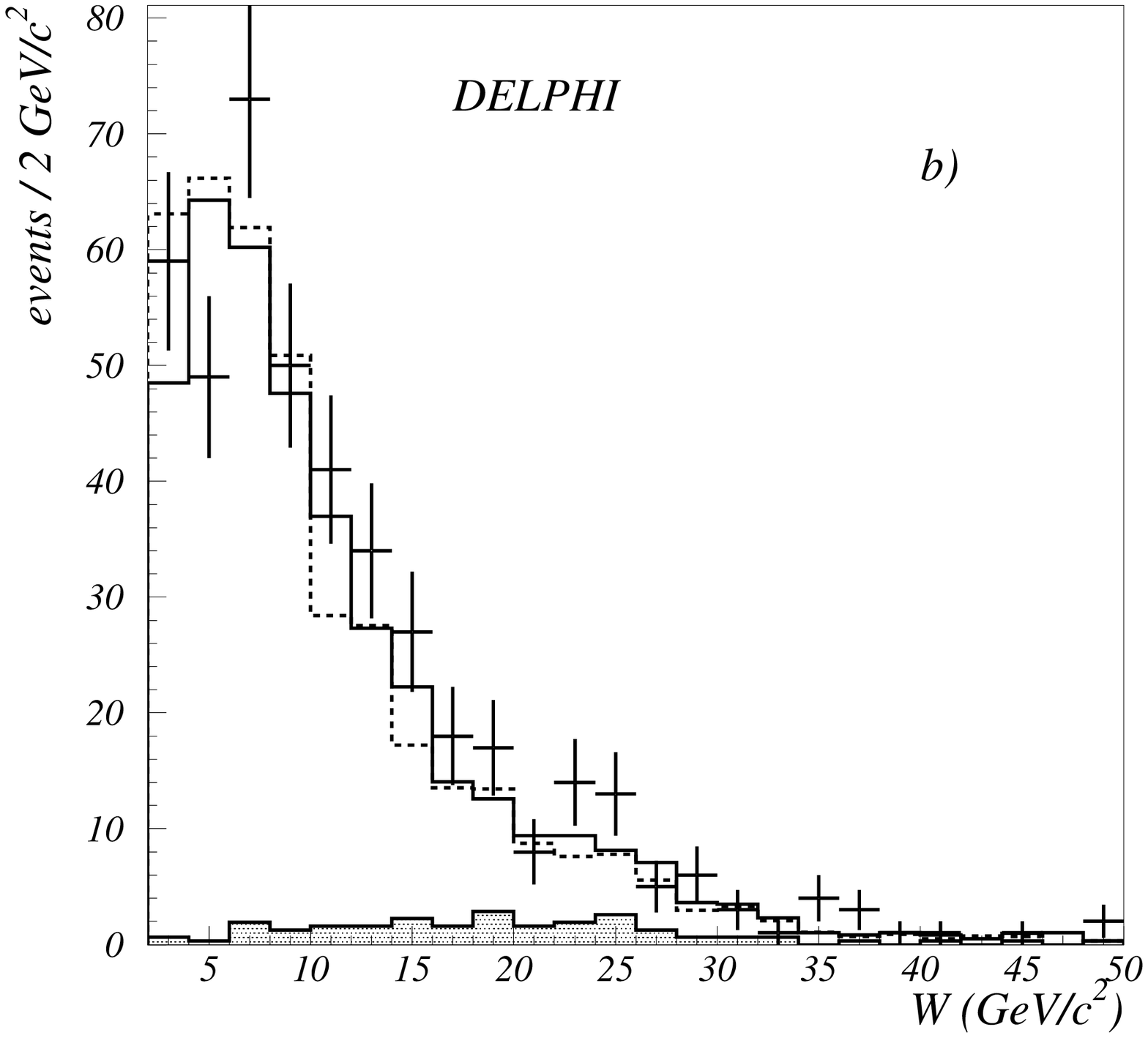,width=8.5cm}}
\end{minipage}
\end{tabular} \\ \\ \\

\begin{tabular}{cc}
\begin{minipage}[h]{9cm}
\vspace*{1cm}
\hspace*{-2.cm}
\mbox{\epsfig{figure=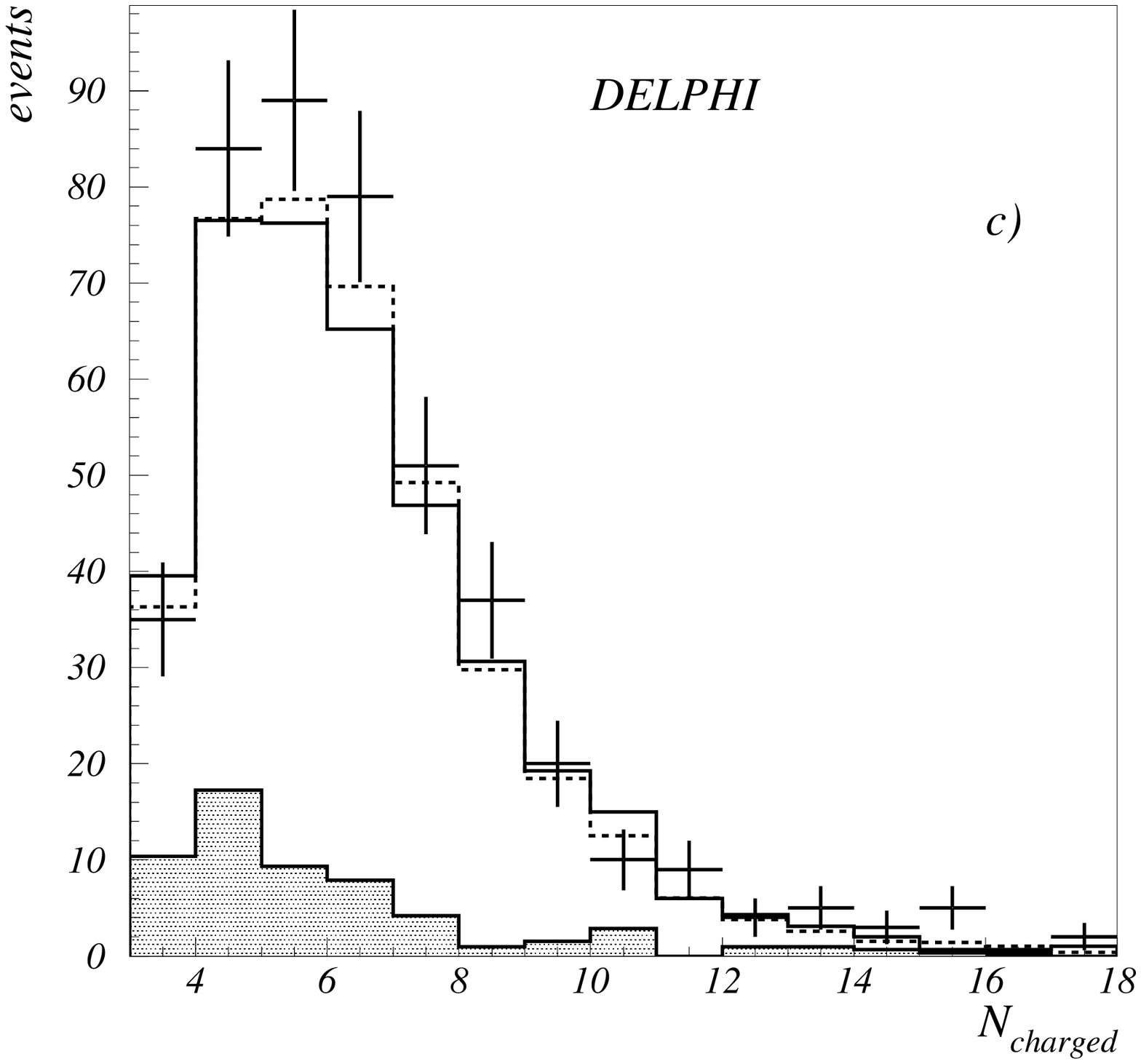,width=8.5cm}}
\end{minipage}
\begin{minipage}[h]{9cm}
\vspace*{1cm}
\hspace*{-2.5cm}
\mbox{\epsfig{figure=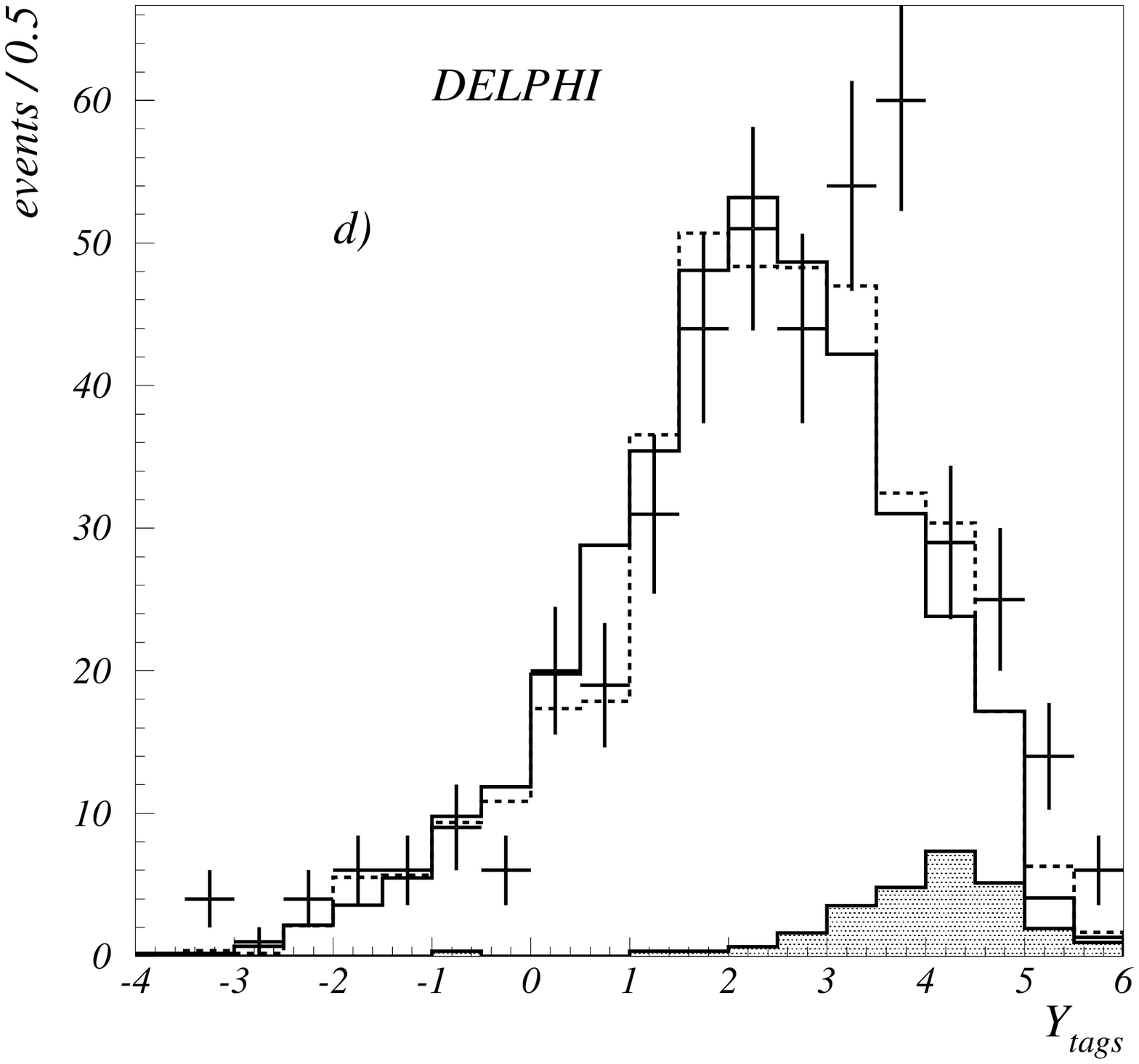,width=8.5cm}}
\end{minipage}
\end{tabular}
\caption[]{Distributions of $Q^2_i$ (a), invariant mass $W_{had}$ (b),
charged multiplicity $N_{charged}$ (c), and $Y$ calculated from the tagged 
particles' 4-momenta (d). The data are shown with error bars.  
The solid and dashed histograms correspond to the sum
of \agg simulated events obtained with the PYTHIA and TWOGAM generators,
respectively, and include the total background, shown hatched.}
\label{fig:fig6}
\end{figure}

 The data distributions for the photon virtualities, $Q^2_i$ (two entries 
per event), the invariant mass of the hadron system calculated with the 
charged particles' 4-momenta, $W_{had}$, the charged particles'
multiplicity and the $Y$ variable calculated with $\wgg$ are compared 
with the Monte Carlo simulation in figure \ref{fig:fig6}. 
The data are represented with error bars. The solid and dashed histograms correspond 
to the sum of \agg simulated events obtained with the PYTHIA and TWOGAM generators,
respectively, and various background sources. The hatched histograms show the
estimated background contamination. Both \agg models agree 
reasonably well with the data. The excess of the data over Monte Carlo
(for low $Q^2$, large $W$) already indicates that the QPM term is insufficient.
\par
\begin{figure}[ht!]
\begin{center}
\mbox{\epsfig{figure=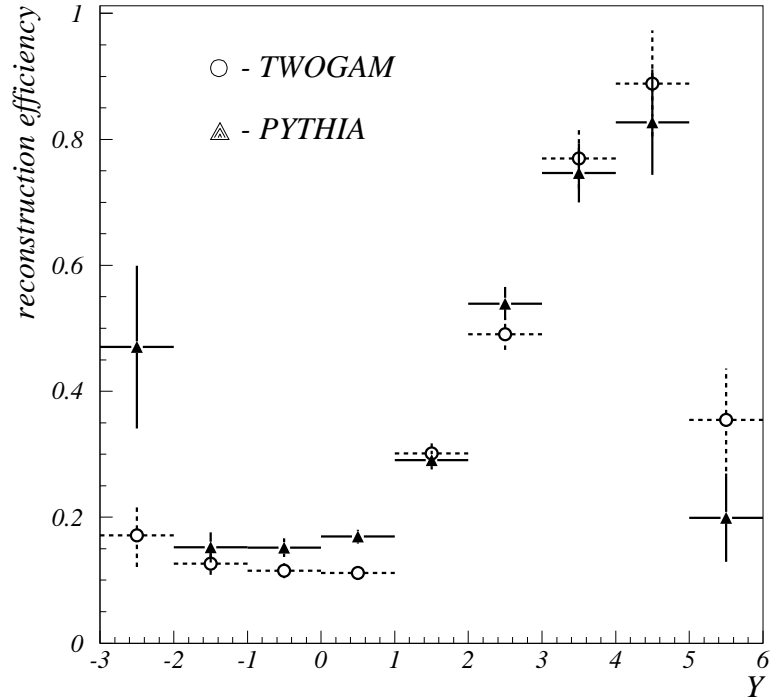,width=11cm}}
\end{center}
\vspace*{-1.5cm}
\caption[]{Reconstruction efficiency as a function of $Y$. The results of the 
calculations based on TWOGAM and PYTHIA generators are shown.}
\label{fig:fig7}
\end{figure}

 The calculations of the detector acceptance and efficiency have been done 
for both \agg models and are shown in figure \ref{fig:fig7}.
The detection efficiencies express slightly different behaviour -
the TWOGAM values are larger than the PYTHIA ones for high values of the $Y$ variable, 
while for small $Y$ values the behaviour is the opposite.
The decrease of the efficiencies for $Y$ above 4 is due to the selection 
criteria. 

\subsection{Results}
\par
\begin{figure}[ht!]
\vspace*{1cm}
\begin{center}
\mbox{\epsfig{figure=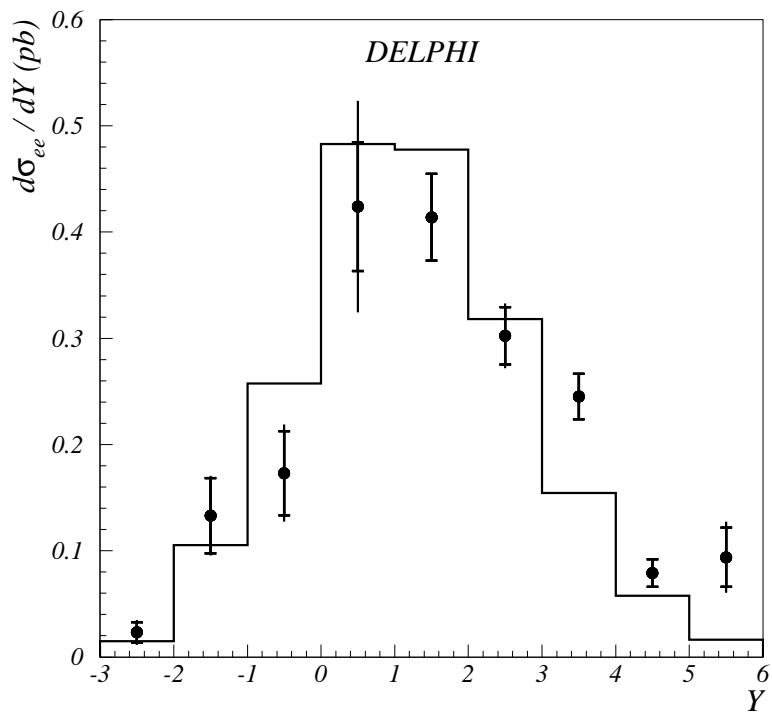,width=11cm}}
\end{center}
\vspace*{-1.5cm}
\caption[]{Differential cross-section for the reaction \ggeeha.
The dashed histogram corresponds to the average of the TWOGAM  and PYTHIA 
predictions. The data are shown with error bars: the total error bars indicate the 
sum in quadrature of the statistical (inner error bars) and of the 
systematic uncertainties.}
\label{fig:fig8}
\end{figure}

 The background subtracted data are corrected for detector effects
using the two models, 
and the measured differential cross-sections \dsee are shown
in figure \ref{fig:fig8} together with the average expectation of 
the two event generators used. 
The uncertainty due to the migration of events
caused by the finite $Q^2$ resolution (the relative uncertainty is around 0.08)
is found to be small in comparison with the statistical uncertainty of the measurement.
Note that, irrespective of the model, the data indicate a somewhat larger cross-section 
compared to expectations for high values of $Y$
corresponding to large invariant masses of the $\gamma\gamma$ system.
The total cross-section \see of the \ggeeha
interactions, within the phase space limited by the criteria $Q^2_i$ 
between 10 GeV$^2$ and 200 GeV$^2$, and $W_{had}$ above 2 GeV/${\it c}^2$,
is measured to be $(2.09\pm0.17)$ pb using the corrections for detector
effects based on TWOGAM and $(1.86\pm0.14)$ pb for the corrections based 
on PYTHIA. The statistical and systematic (see later) uncertainties are added in
quadrature. The expectation of the quark-parton model is $(1.81\pm0.02)$ pb
as obtained with TWOGAM.

\par The \gghadt interactions are expected to be sensitive to
multiple gluon exchange (fig.\ref{fig:fig1}). 
The multigluon ladder is described by the BFKL equation \cite{bfkl}, which
predicts a growth of the cross-section at large $Y$.
Note that the BFKL calculations are valid provided
$\wgg^2 \gg Q^2_i$ (the variable $Y$ should be larger than 2) and
$\mid\ln(Q^2_1/Q^2_2)\mid < 1$ (to maintain the photon virtualities
approximately equal). The application of this latter condition has the effect
of reducing the data sample by about 37\%. It has to be mentioned however 
that the migration of events around the chosen cut at unity does not 
introduce an appreciable systematic uncertainty.
According to the Monte Carlo simulation, around 3\% of the selected 
sample had the true value of the logarithm above unity and pass
the cut due to the $Q^2$ resolution. Approximately the same percentage
migrates inversely.

\par The experimental conditions of the present study ($Q^2_i 
\gg m_e^2$ and the symmetry requirement for tagged particle detection)
permit the relation between \see and \sgg,  
which initially reads \cite{bud}
(the interference terms are omitted):
\begin{center}
$\sigma_{ee} = \sum_{i,j=T,L} L_{ij}\sigma_{ij}$, 
\end{center}
to be simplified to a relation involving an effective cross-section \sgg, 
\begin{center}
$\sigma_{ee} = L_{TT}\sigma_{\gamma^*\gamma^*}$ with $
\sigma_{\gamma^*\gamma^*} = \sigma_{TT} + 2\epsilon\sigma_{LT} + 
\epsilon^2\sigma_{LL}$,
\end{center}
where $L_{TT}$ is the flux 
of the transversely polarized photons 
calculable in QED, $\epsilon$ is around 0.94,
$\sigma_{LT}\simeq 0.2\sigma_{TT}$ and $\sigma_{LL}\simeq 0.05\sigma_{TT}$
\cite{brod}.
The TWOGAM event generator including QED radiative 
corrections has been used to calculate $L_{TT}$: 
it uses the decomposition of the cross-section for different photon 
helicities \cite{bud}. The limits on $Q^2_i$, $\mid\ln(Q^2_1/Q^2_2)\mid$ 
and $W_{had}$ are the same as described above.
The differential cross-sections \dsee, both for data and MC, and \dsgg 
are all presented in Table 1. The selection efficiency is calculated using
the mean of the results using TWOGAM and PYTHIA. The difference between the results
using these two generators is used to calculate the systematic error
due to modeling and is in included in the quoted systematic uncertainties.
The flux binned over $Y$ is presented as well.
The absolute uncertainties on the $dL_{TT}/dY$ calculations are of the order of 
$0.002\times10^{-3}$.

\begin{center}
\small
\begin{tabular}{|c|c|c|c|c|} \hline
$Y $ & \dseed & \dseemc& $dL_{TT}/dY$  & \dsgg \\
     & (pb) & (pb) & ($\times 10^3$)      & (nb) \\ \hline
& \multicolumn{2}{c} {no $\ln(Q^2_1/Q^2_2)$ cut} \vline & 
  \multicolumn{2}{c} { $-1 < \ln(Q^2_1/Q^2_2) < 1 $} \vline \\ \hline
($-$3)-($-$2) &0.02\pv0.01(stat)\pv0.01(syst) &0.02\pv0.01 &0.060 & 0.20\pv0.11(stat)\pv0.06(syst) \\ \hline
($-$2)-($-$1) &0.13\pv0.04(stat)\pv0.01(syst) &0.11\pv0.02 &0.114 & 0.56\pv0.18(stat)\pv0.15(syst) \\ \hline
($-$1)-0    &0.17\pv0.04(stat)\pv0.02(syst) &0.26\pv0.03 &0.113 & 0.89\pv0.23(stat)\pv0.21(syst) \\ \hline
0-1       &0.42\pv0.06(stat)\pv0.08(syst) &0.48\pv0.01 &0.090 & 2.50\pv0.49(stat)\pv0.58(syst) \\ \hline
1-2       &0.41\pv0.04(stat)\pv0.01(syst) &0.48\pv0.01 &0.082 & 3.56\pv0.42(stat)\pv0.08(syst) \\ \hline
2-3       &0.30\pv0.03(stat)\pv0.02(syst) &0.32\pv0.01 &0.070 & 3.00\pv0.33(stat)\pv0.19(syst) \\ \hline
3-4       &0.25\pv0.02(stat)\pv0.01(syst) &0.15\pv0.01 &0.054 & 2.83\pv0.32(stat)\pv0.02(syst) \\ \hline
4-5       &0.08\pv0.01(stat)\pv0.01(syst) &0.06\pv0.01 &0.034 & 1.47\pv0.29(stat)\pv0.07(syst) \\ \hline
5-6       &0.09\pv0.03(stat)\pv0.02(syst) &0.02\pv0.01 &0.015 & 3.51\pv1.33(stat)\pv0.60(syst) \\ \hline
\end{tabular}
\normalsize
\end{center}
Table 1: The measured and expected differential cross-sections 
\dseed of the reaction \ggeeha, the calculated photon flux $L_{TT}$ 
including the radiative corrections \cite{twogam}, and 
the measured cross-section \dsgg of the process \gghadt are shown 
as a function of the variable $Y$.

\par
\begin{figure}[ht!]
\begin{center}
\hspace*{-2.cm}
\epsfig{file=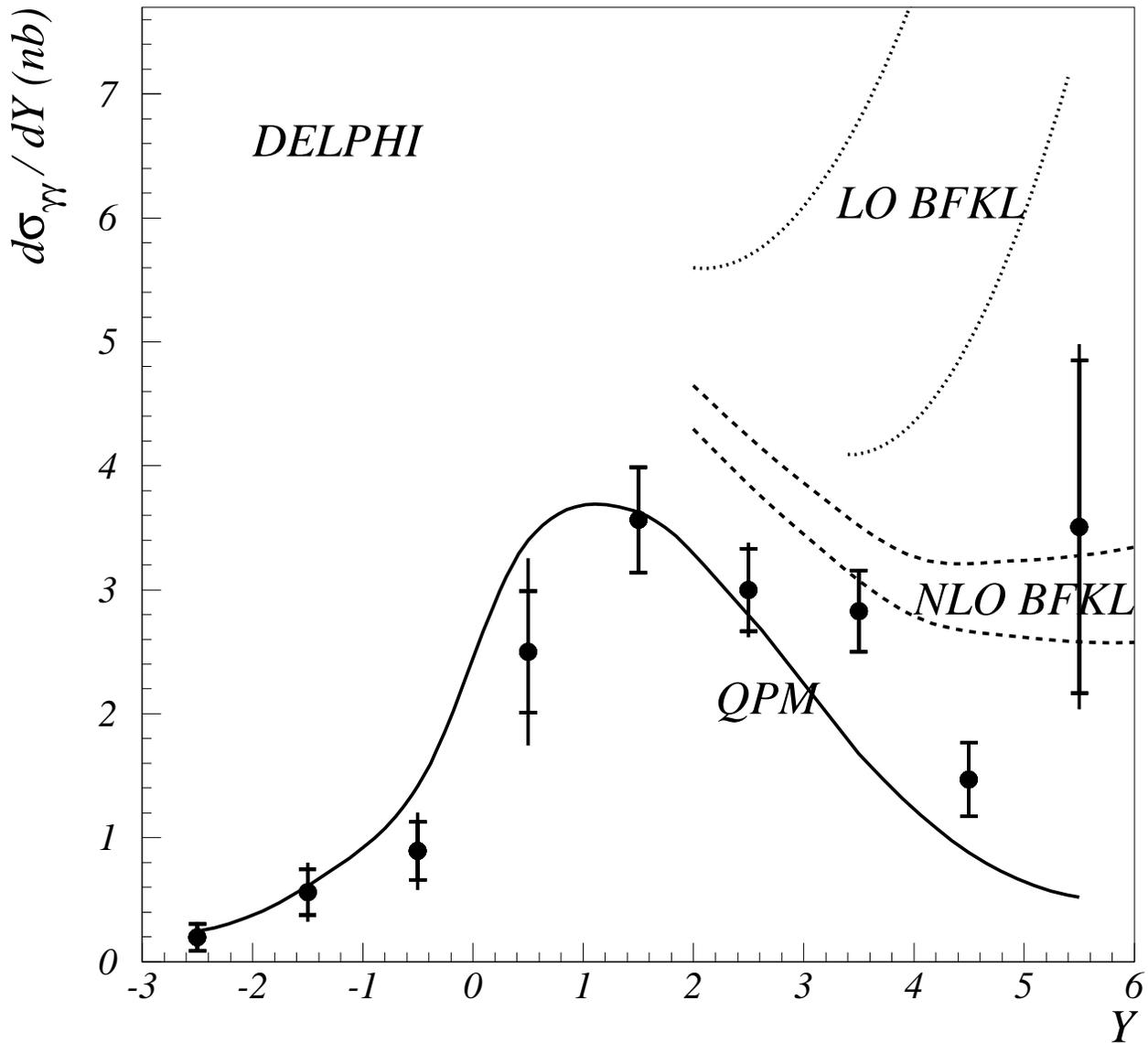,width=18.0cm}
\end{center}
\vspace*{-2cm}
\caption[]{The differential cross-section for the reaction \gghadt. 
The data are shown with error bars: the total error bars indicate the 
sum in quadrature of the statistical (inner error bars) and of the 
systematic uncertainties.  
The solid curve corresponds to the expectation of the quark-parton 
model (QPM, quark-box diagram, figure 1). The two dotted lines represent 
the BFKL calculations in the leading order \cite{ewerz}. The 
next-to-leading order calculations \cite{kim} are shown by the two dashed 
curves in the middle.
The two curves for the BFKL calculations correspond to the Regge scale 
parameter changing between $Q^2$ (upper line) and 4$Q^2$ (lower one).
The QPM contribution is added to both the LO and the NLO BFKL expectations.
}
\label{fig:fig9}
\end{figure}

 The measured differential cross-section \dsgg is shown in figure 
\ref{fig:fig9}.
The systematic uncertainties are dominated by the difference between
the results obtained with TWOGAM and PYTHIA Monte Carlo generators.
The other systematic uncertainties, calculated by varying the selection 
criteria, the $Q^2$ domain, etc. represent between 17\% and 29\% of 
the statistical uncertainties.
The predictions of the QPM and of BFKL calculations, both in LO \cite{ewerz} and NLO \cite{kim}, 
are also shown in figure \ref{fig:fig9}. 
The two curves for the BFKL calculations correspond to the Regge scale 
parameter $s_0$, which defines the start of the asymptotic regime, 
equal to $Q^2$ or 4$Q^2$. The LO calculations are more affected by the 
uncertainty coming from the choice of the scale parameter than the NLO ones. 
Note that the BFKL calculations are weighted 
over a number of $Q^2$ bins and that therefore the running of $\alpha_s$ is also
included. The data lie in any case much lower than the BFKL cross-sections 
calculated in leading-order. On the other hand, the data are closer to
NLO predictions, since the expected growth of the
gluon exchange contribution (BFKL) is much weaker and appears mainly 
for $Y$ values larger than 4. Below this value the cross-section
is dominated by the decrease of the QPM contribution. Unfortunately the 
LEP energy and the present statistics are not sufficient to study in 
detail the region at large $Y$, where BFKL is expected to dominate.  

\section{Conclusions}
\par Double-tagged \agg interactions have been studied with the DELPHI data 
taken at $e^+e^-$ centre-of-mass energies from 189 GeV to 209 GeV and
corresponding to an integrated luminosity of 550 pb$^{-1}$.
For virtualities, $Q^2$, of both photons between 10 GeV$^2$ and 200 GeV$^2$ 
and final state invariant mass $W$ above 2 GeV/${\it c}^2$, the cross-section of 
the process \ggeemm is measured to be (1.38$\pm0.12(stat)\pm0.06(syst))$ pb,
to be compared with the expectation of (1.36$\pm0.01)$ pb for the QED 
calculations including radiative corrections to the photon flux.
The cross-section \see of the \ggeeha interactions is measured to be
$(2.09\pm0.09(stat) \pm0.19(syst))$ pb with the corrections for detector 
effects based on the TWOGAM event generator \cite{twogam}.
The differential cross-section \dsgg of the \gghadt interactions is measured
and is compared with the predictions based on LO and NLO BFKL calculations.
The leading order calculations clearly
disagree with the data while the next-to-leading order predictions are found
to be more consistent with the data, although the LEP energy is not
sufficient to see a sizable effect due to the BFKL type contribution.
The DELPHI data are in agreement with the results of the other LEP experiments
\cite{others}.

\subsection*{Acknowledgements}
\vskip 3 mm
 We are greatly indebted to our technical 
collaborators, to the members of the CERN-SL Division for the excellent 
performance of the LEP collider, and to the funding agencies for their
support in building and operating the DELPHI detector.\\
We acknowledge in particular the support of \\
Austrian Federal Ministry of Education, Science and Culture,
GZ 616.364/2-III/2a/98, \\
FNRS--FWO, Flanders Institute to encourage scientific and technological 
research in the industry (IWT), Belgium,  \\
FINEP, CNPq, CAPES, FUJB and FAPERJ, Brazil, \\
Czech Ministry of Industry and Trade, GA CR 202/99/1362,\\
Commission of the European Communities (DG XII), \\
Direction des Sciences de la Mati$\grave{\mbox{\rm e}}$re, CEA, France, \\
Bundesministerium f$\ddot{\mbox{\rm u}}$r Bildung, Wissenschaft, Forschung 
und Technologie, Germany,\\
General Secretariat for Research and Technology, Greece, \\
National Science Foundation (NWO) and Foundation for Research on Matter (FOM),
The Netherlands, \\
Norwegian Research Council,  \\
State Committee for Scientific Research, Poland, SPUB-M/CERN/PO3/DZ296/2000,
SPUB-M/CERN/PO3/DZ297/2000, 2P03B 104 19 and 2P03B 69 23(2002-2004)\\
FCT - Funda\c{c}\~ao para a Ci\^encia e Tecnologia, Portugal, \\
Vedecka grantova agentura MS SR, Slovakia, Nr. 95/5195/134, \\
Ministry of Science and Technology of the Republic of Slovenia, \\
CICYT, Spain, AEN99-0950 and AEN99-0761,  \\
The Swedish Research Council,      \\
Particle Physics and Astronomy Research Council, UK, \\
Department of Energy, USA, DE-FG02-01ER41155. \\
EEC RTN contract HPRN-CT-00292-2002. \\


\clearpage

\vspace*{2cm}
\vspace*{1cm}
\vspace*{-2cm}


\end{document}